\documentclass[11pt]{llncs}
\usepackage{graphicx}
\usepackage{dcolumn}
\usepackage{bm}
\usepackage{amssymb,amsmath}

\begin{document}%

\title{Game Theoretical Interactions of Moving Agents}
\author{Wenjian Yu and Dirk Helbing}
\institute{ETH Zurich, UNO D11, Universit\"atstr. 41, 8092 Zurich,
Switzerland} \maketitle
\begin{abstract}
Game theory has been one of the most successful quantitative
concepts to describe social interactions, their strategical aspects,
and outcomes. Among the payoff matrix quantifying the result of a
social interaction, the interaction conditions have been varied,
such as the number of repeated interactions, the number of
interaction partners, the possibility to punish defective behavior
etc. While an extension to spatial interactions has been considered
early on such as in the ``game of life'', recent studies have
focussed on effects of the structure of social interaction networks.

However, the possibility of individuals to move and, thereby, evade
areas with a high level of defection, and to seek areas with a high
level of cooperation, has not been fully explored so far. This
contribution presents a model combining game theoretical
interactions with success-driven motion in space, and studies the
consequences that this may have for the degree of cooperation and
the spatio-temporal dynamics in the population. It is demonstrated
that the combination of game theoretical interactions with motion
gives rise to many self-organized behavioral patterns on an
aggregate level, which can explain a variety of empirically observed
social behaviors.
\end{abstract}

\section{Introduction}

Macroscopic outcomes in a social system resulting from interactions
between individuals can be quite different from anyone's intent. For
instance, empirical investigations \cite{soc_context_seg} have shown
that most colored people prefer multi-racial neighborhoods, and many
white people find a certain fraction of other races in their
neighborhood acceptable. So one could think that integrated
neighborhoods should be widely observed, but empirically this is not
true. One rather finds segregated neighborhoods, i.e. separate urban
quarters, which also applies to people with different social and
economic backgrounds.

This problem is scientifically addressed by mainly two streams of
segregation theory \cite{ethnic_model}: the urban ecological
``social distance" tradition in sociology
\cite{social_dist0,social_dist1} and the ``individual preferences"
tradition in economics \cite{dyna_model_segre,micro_macro}. The main
idea of ``social distance" theory is that the differences in culture
and interests between social groups are reflected by a separation of
their residential areas.

Yet, the role of social distance and individual preferences is
questioned by studies of the American urban housing market, which
suggest that racial discrimination and prejudices are the primary
factors of residential segregation and concentration of poverty
\cite{seg_poverty}. There are three stages in a housing market
transaction: first, information about available housing units,
second, terms and conditions of sales and  financing assistance, and
third, the access to units other than the advertised unit
\cite{closed_doors}. In each stage, the housing agent may behave in
a discriminatory way, e.g. withhold information from customers and
discourage them. Therefore, the access of minority customers to
housing is severely constrained, while the theory of
preference-based dynamics assumes that people can relocate freely
according to their own preferences, which fails to reflect the real
relocation dynamics.

Recent studies \cite{ethnic_model,ethnic_theo} point out that the
effect of discrimination in racial residential segregation was
important in the past, but nowadays, the nature and magnitude of
housing discrimination has changed. Minority households who seek to
move into integrated or predominantly white areas usually will be
able to do so. But the integration of ethnic groups is not widely
observed and quite unstable. This is partly because, when people
move to a neighborhood, where they constitute an ethnic minority,
the previous inhabitants may choose to leave, while some people may
be reluctant to enter integrated neighborhoods in which minorities
are increasing in number, e.g. because of a decrease in the housing
prices. Such migration dynamics, based on seeking preferred
neighborhoods, does not necessarily presuppose the discrimination of
other people, as we will shortly see.

Assuming that individuals have just a slight preference for
neighborhoods, in which the same ethnic background prevails,
Schelling has reproduced residential segregation by a simple model
\cite{dyna_model_segre,micro_macro}. Imagine that two groups of
people are distributed over a one-dimensional lattice, and assume
that everyone defines his/her relevant neighborhood by the four
nearest neighbors on either side of him/her. Moreover, assume that
each individual prefers at least four of his/her eight nearest
neighbors to belong to same ethnic group as he/she does. Considering
himself/herself, this implies a small majority of five out nine
people in the considered neighborhood. If the condition is not met,
he or she is assumed to move to the nearest site/cell that satisfies
his or her desire. Hence, the model does not assume optimizing
location choice, just satisficing behavior. Nevertheless, it
produces segregation patterns.

\subsection{Migration, Game Theory, and Cooperation}

In game theoretical terms, this movement to a more favorable
neighborhood could be reflected by a higher payoff to persons who
moved, called ``migrants" in the following. Migratory behavior
aiming at higher payoffs is called ``success-driven motion"
\cite{success_driven,opt_slef_org,mig_game}. As we will show in
later sections, success-driven motion can reproduce residential
segregation and some other observed phenomena of population dynamics
as well, like population succession (i.e. cycles of changing
habitation) \cite{pop_success}. We will also study how a change of
the spatial population structure can affect the level and evolution
of cooperation in a population.

It could be thought that natural selection, implying competition
between individuals, would result in selfish rather than cooperative
behavior \cite{five_rules_for_the_evolution_of_cooperation}.
Nevertheless, cooperation is widely observed also in competitive
settings, from bacteria, over animals to humans
\cite{the_evolution_of_cooperation,coop_comp_pathogenic_bacteria}.
Game theory
\cite{neumann_game,game_fuden,game_andreas,irregular_latt} has been
regarded as a powerful framework to investigate this problem, as it
can be used to quantify the interactions between individuals.

The mathematical description of tendencially selfish behavior is
often based on the prisoner's dilemma game, in which the ``reward"
$R$ represents the payoff for mutual cooperation, while the payoff
for defection (cheating) on both sides is reflected by the
``punishment" $P$. Unilateral cooperation will incur the so-called
``sucker's payoff" $S$, while the defector gains T, the
``temptation". Given the inequalities $T>R>P>S$ and $2R>T+S$,
defection is tempting and cooperation is risky, so that defection is
expected to be dominating, although mutual cooperation improves the
average payoff.

In a well-mixed population, where an individual interacts with all
the others, a defector's payoff is always higher than the average
payoff, which leads to a prosperity of defectors. In an evolutionary
setting as it is described by the replicator dynamics
\cite{replicator_dynamics}, only those strategies that generate
above-average payoffs have the chance to spread. Therefore, from an
evolutionary perspective, there will be no cooperators in the end,
which contradicts the observed cooperation in reality.

A variety of mechanisms has been proposed to explain the
considerable level of cooperation observed under certain
circumstances. These include kin selection, direct reciprocity,
indirect reciprocity, group selection and network reciprocity
\cite{five_rules_for_the_evolution_of_cooperation}. In particular,
it is interesting that the spatial population structure can
significantly change the level of cooperation in a population
\cite{spatial_chaos,spatial_structure_often_inhibits_the_evolition}.
While there could be a co-evolution of social structure and
cooperation due to migratory behavior, this subject has not been
well studied in the past.

\subsection{Migration and Population Succession}
Population succession plays an important role in the change of the
population structure in urban areas. The term succession is viewed
by ecologists as an orderly sequence of changes in the composition
or structure of a biotic community \cite{human_ecology}. Later on,
it was also used to describe the change in neighborhood and urban
land use \cite{neighbor_change}. For example, Cressey
\cite{pop_success} analyzed the movements of immigrants in Chicago
from 1898 to 1930 by census tracts and provided a detailed picture
of the succession processes of specific cultural groups.
Interestingly, these processes show a cycle of invasion, conflict,
recession, and reorganization \cite{weidlich0}.

The ``invasion" can be initiated by a few ``pioneers", who manage to
enter a new area dominated by ``older" residents. For example, some
rich colored people may move to a white-only neighborhood, where the
housing price is relatively high. In this stage, discrimination by a
housing agency could prevent the integration of cultural groups by
constraining information to minority groups \cite{closed_doors}. But
nowadays, housing discrimination is illegal in many countries, and
the hostility toward minority group members has declined. Therefore,
the possibility of successful entrance for those ``pioneers" is much
larger.

Following the initial ``invaders", invasions can happen on a much
larger scale, in particular if housing prices go down. Then,
previous residents, who want to sell their houses at a good price,
are eventually replaced. People may even move without reasons, just
following the trend of what the others do, which is a kind of
collective herding behavior \cite{coll_behav}. Accompanying this
replacement process, the changing composition of the population may
cause conflicts due to incompatible cultural values and prejudices
\cite{value_drop}.


Eventually, the population composition is turned over. Although the
new group can dominate the residential area, it takes time to
reorganize the environment such as churches and to create it
according to their own requirements. The cycle can repeat as the
previous ``invaders" become established residents. Therefore, there
is a typical life circle of a neighborhood \cite{neighbor_change},
which involves five stages: development, transition, downgrading,
thinning out, and renewal. If the facilities in a neighborhood are
running down, residents start to seek new places and abandon the
previous one. In this case, there is usually little conflict. In the
subsequent renewal stage, new people move in at low prices. By
renovating the houses, the quarter eventually revives.

\subsection{Co-evoultion of Social Structure and Cooperation}

In order to model such intentional movement of individuals, spatial
effects must be explicitly considered. Spatial games based on
lattices \cite{spatial_chaos} would allow one to study such effects.
In conventional spatial games, individuals are uniformly distributed
in the simulation area, and change or repeat their strategies,
following a certain updating rule. For instance, an individual is
assumed to unconditionally adopt the most successful strategy within
the neighborhood. On the one hand, this creates typical
spatio-temporal pattern. On the other hand, spatial structures play
an important role in the maintenance of cooperation. In the
iterative prisoner's dilemma, for example, clusters of cooperators
are beneficial to cooperators \cite{mig_game}, while the evolving
spatial structure in the snowdrift game
\cite{spatial_structure_often_inhibits_the_evolition}, may inhibit
cooperation. Conventional spatial games, however, neglect the
possibility of individuals to move or migrate, although mobility is
a well-known fact of daily social interactions. We will see that the
movement of individuals, e.g. population succession, can change the
spatial structure of a population significantly, and influence the
evolution of cooperation dramatically.

The above mentioned social processes that are related to the
migratory behavior of people, can be well integrated into the
framework of spatial games. Focusing on success-driven motion, we
will in the following, study the combination of migration, strategic
interactions and learning (specifically imitation). Numerical
simulations show that success-driven motion is a mechanism that can
promote cooperation via self-organized social structures. Moreover,
very surprisingly, a certain degree of fluctuations (``noise") can
even enhance the level of cooperation.

\section{Spatial Games with Mobility}

\subsection{Classification}
Spatial games with mobility (``mobility games'') could be classified
as follows:
\begin{enumerate}
\item Mobility may take place in physical (geographic) space with one, two or three dimensions,
or it may occur in abstract space (e.g. opinion space). Rather than
a regular (grid) space, one may also have a (fixed or dynamically
changing) network structure (e.g. friendship network).
\item One may distinguish between games with continuous and with discrete motion. The first ones may be considered as particular cases of differential games \cite{diff_game} and shall be called ``motion games''. The second ones will be named ``migration games" and can be implemented, for example, in terms of cellular automata (particularly, if a parallel update is performed).
\item The mobility game may be deterministic or ``noisy'', i.e. influenced by fluctuations. Such ``noise'' may be introduced by stochastic update rules (e.g. a random sequential update or the Fermi rule \cite{spatial_structure_often_inhibits_the_evolition}.
\item Multiple occupation of a certain location may be possible, allowing for the agglomeration of individuals, or it may be prohibited. In the latter case, spatial exclusion requires spatial capacity constraints and/or the respect or protection of some ``private territory''.
\item In mobility games with spatial exclusion, empty sites are needed. Therefore, the density of free locations (``sites'') is a relevant model parameter.
\item The frequency, rate, or speed of mobility may be relevant as well and should be compared with other time scales, such as the frequency of strategy changes, or the average lifetime of a spatial cluster of individuals. Depending on the specification of the game, after a transient time one may find everything from chaotic patterns \cite{spatial_chaos} upto frozen patterns \cite{dis_env,mob_dec}. This may be expressed by a viscosity parameter \cite{mob_visc}.
\item Mobility may be random or directed (e.g. ``success-driven''). Directed mobility may depend on the expected payoff in a certain time point (iteration), or it may depend on the cumulative payoff (i.e. the accumulation of wealth by individuals).
\item If age, birth and death are considered, one speaks of ``demographic games'' \cite{ca0}.
\end{enumerate}
In the following, we will primarily focus on games with
success-driven migration in two-dimensional geographic space.
Furthermore, we will assume simple strategies (such as all-cooperate
or all-defect), no memory and no forecasting abilities.

\subsection{Individual Decision Making and Migration}
\begin{figure}[!htbp]
\begin{center}
\includegraphics[width=5cm, angle = 0]{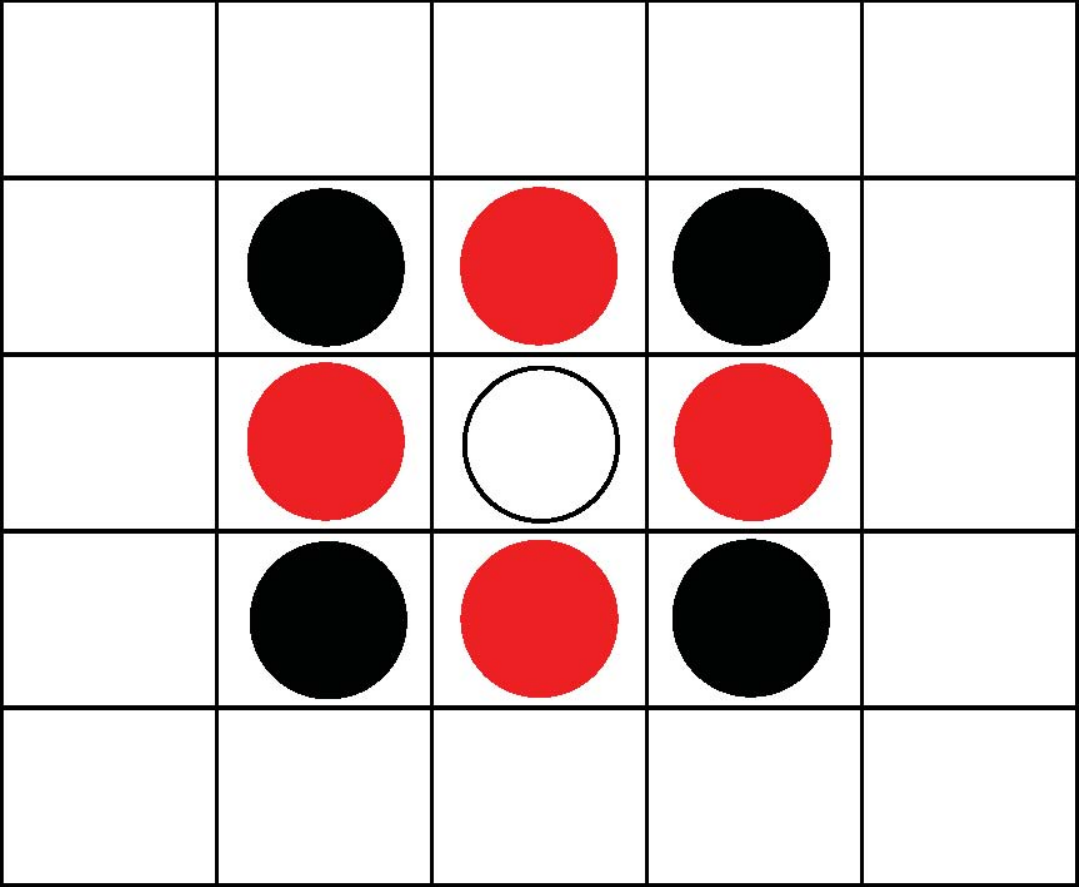}\,
\caption[]{The focal individual $I$ is represented by the empty
circle. It can, for example, interact with the $k = 4$ nearest
individuals (red) or with $k = 8$ neighbors, which includes four
next-nearest neighbors (black).} \label{fig_MooreNeighborhood}
\end{center}
\end{figure}

In ``migration games", each individual $I$ is located at the
position $\vec{X_{I}}$ of a discrete grid and applies a certain
strategy $i=i(I,\vec{X},t)$ when interacting with individuals $J$ at
locations $\vec{X_{J}}$ in the neighborhood ${\cal N} = {\cal
N}(\vec{X_{I}})$, who apply strategies $j=j(J,\vec{X'},t)$ (see
Fig.~\ref{fig_MooreNeighborhood}). The interactions at time $t$ are
quantified by the payoffs $P_{ij}$. The overall payoff for
individual $I$ resulting from interactions with all individuals $J$
in the neighborhood ${\cal N}(\vec{X_{I}})$ at time $t$ is

\begin{eqnarray}
P_{I}(t) = P_{I}(\vec{X_{I}},t)=\sum_{J:\vec{X_{J}}\in {\cal
N}(\vec{X_{I}})}P_{ij}(J).
\end{eqnarray}

\par

\begin{figure}[!htbp]
\begin{center}
\includegraphics[width=10cm, angle = 0]{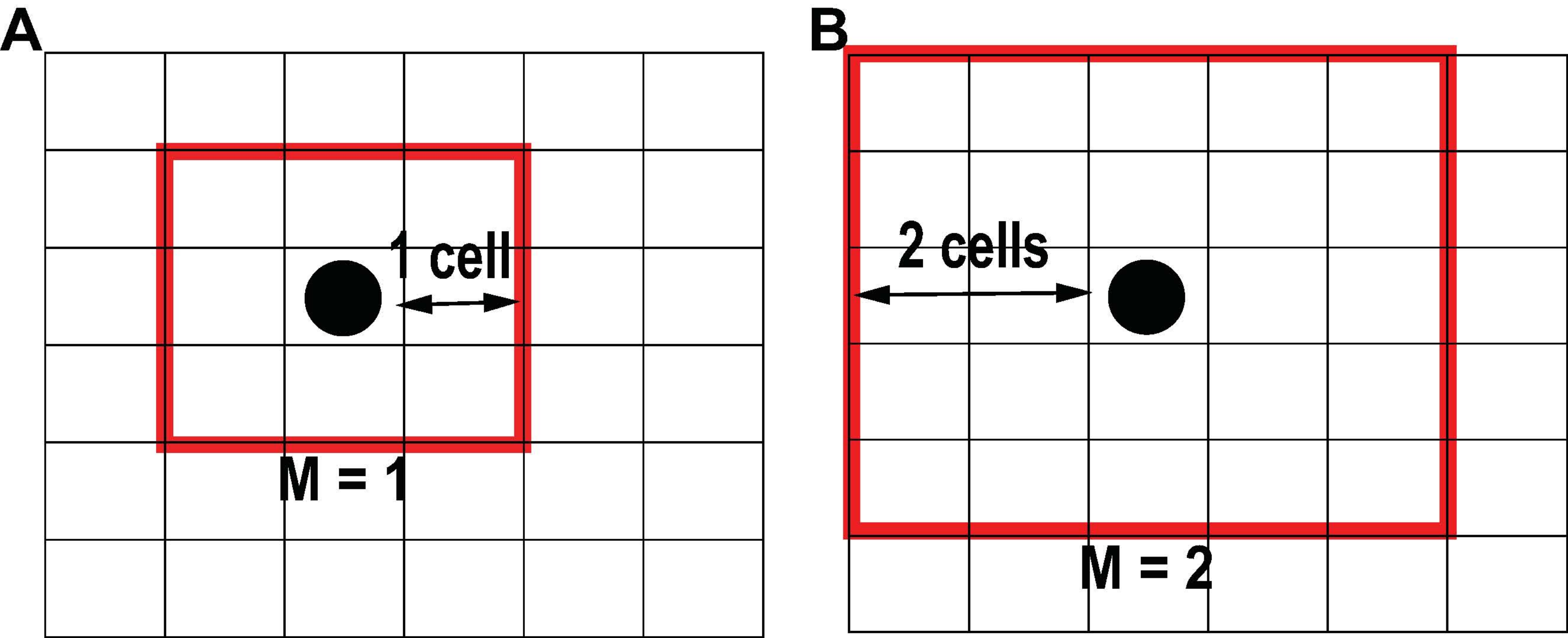}\,
\caption[]{\textbf{(A)} When $M=1$, the focal individual (black
circle) can migrate within the red square (Moore neighborhood of
range 1). \textbf{(B)} Here, the migration neighborhood is chosen as
Moore neighborhood of range $M=2$.} \label{fig_migration_range}
\end{center}
\end{figure}

We assume that all individuals prefer places that can provide higher
payoff. However, their movements are restricted by their mobilities
and the number of free locations, as one can only move to empty
places. In our model, the mobility is reflected by the migration
range $M$, which could be assumed constant (see
Fig.~\ref{fig_migration_range}) or as a function of cumulative
payoffs

\begin{eqnarray}
C_{I}(t) = \sum_{t'=0}^{t}[P_{I}(t') - c_{I}(t')].
\end{eqnarray}

Here, the cost of each movement $c_{I}(t)$ may be specified
proportionally to the distance $d_{I}(t)\leq M$ moved:
\begin{eqnarray}
c_{I}(t) = \beta~d_{I}(t).
\end{eqnarray}
$\beta$ is a constant.

When the migration range $M$ is restricted by the cumulative payoff,
one may set
\begin{eqnarray}
M(t) = \lfloor\alpha~C_{I}(t)\rfloor,
\end{eqnarray}
where $\lfloor~\rfloor$ rounds down to the next integer $\leq \alpha
C_{I}(t)$ and $\alpha$ is a constant, which is set to
$\frac{1}{\beta}$ in the following.

\par

The expected payoff for individual $I$ applying strategy $i$ at a
free location $\vec{X'}$ within the migration range can be
calculated as
\begin{eqnarray}
P_{I}(\vec{X'},t) = \sum_{J':\vec{X_{J'}}\in {\cal
N(\vec{X'})}}P_{ij'}(J'),
\end{eqnarray}
where $j'=j'(J',\vec{X'},t)$ are the strategies of the individuals
$J'$ located within the neighborhood ${\cal N'} = {\cal
N}(\vec{X'})$ centered at $\vec{X'}$. Here, we assume that an
individual can determine this value by testing a neighborhood $\cal
N'$ by means of some kind of ``fictitious play" with its residents.

All individuals are assumed to maximize the expected payoff by
migrating to the place that promises the highest payoff.
Consequently, the payoff at the new destination for $I$ is expected
to be
\begin{eqnarray}
P_{I}^{e}(t+1) = \max_{\vec{X'}\in {\cal A}}{ P_{I}(\vec{X'},t) },
\end{eqnarray}
where ${\cal A}$ represents the area within the migration range $M$
around $\vec{X_{I}}$.

In our model, individuals decide to move, if the expected payoff at
the new position $P^{e}_{I}(t+1)$ satisfies the short-term
cost-benefit condition
\begin{eqnarray}
P^{e}_{I}(t+1) - P_{I}(t) > c_{I}(t).
\end{eqnarray}

If we assume that the cost of each movement is small, and the new
neighborhood yields a comparatively high cumulative payoff over the
time period the individual stays at that place, then the cost of
movement can be neglected.

Finally, if two or more places promise the same maximum payoff, we
assume that an individual will choose the closest one. If both, the
payoff and distance are the same, then the individual will randomly
choose one of the closest locations in our simulation. Since all the
other individuals also seek new places that can increase their
payoffs, neighborhoods may change quickly, and the resulting payoff
may fall below the expectations before the move (during the
fictitious play step).

\subsection{Learning}
Learning allows individuals to adapt their behaviors in response to
other people's behaviors in an interactive decision making setting.
Basic learning mechanisms are, for example, unconditional imitation,
best reply, and reinforcement learning \cite{indivi_stg}.

Unconditional imitation means that people copy the behaviors of
others. For instance, we can assume that, when $i$ reaches a new
place, it imitates the most successful strategy within the
neighborhood (if it is more successful than the own strategy), or
the most frequently used strategy in past interactions. It should be
underlined, that it is not easy to identify the future strategy of
another individual from its displayed past behavior. However, in the
simplified context of a prisoner's dilemma, where individuals are
assumed to play either all-defect or all-cooperate, one's behavior
reveals the strategy directly. It may nevertheless change before the
next iteration due to learning of the neighbors.

Reinforcement learning is a kind of backward-looking learning
behavior, i.e. people tend to take the actions that yielded high
payoffs in the past. Assuming that an individual has a set of
strategies, among which he or she chooses with a certain
probability, this probability is increased, if the gained payoff
exceeds a certain threshold value (``aspiration level")
\cite{learn_macy0,learn_macy1}.

Best reply is a sophisticated learning model as well. According to
it, people choose their strategies based on the expectation of what
the others will do in a way that maximizes their expected payoff.
The ability to forecast depends on one's belief about the behaviors
of the others. For instance, one can try to determine the
distribution of the neighbors' previous actions, and select the own
strategy, which replies to it in the best way.

For simplicity, players in our model only interact with the $k=4$ nearest neighbors and adapt their strategies by unconditional imitation. Assuming that the updating rules of all the individuals are the same, we can specify a simple migration game as follows:\\
1. An individual moves to a new position that is located inside the migration range $M$ and maximizes the expected payoff.\\
2. The individual imitates the most successful strategy within the
interaction neighborhood ${\cal N}$, if it is more successful than
the own strategy.

Note that changing the sequence leads to a different dynamics, and
the implementation of migration and learning can be varied in many
ways. Therefore, spatial games with mobility promise to be a rich
research field. In the following, we will restrict ourselves to some
of the simplest specifications.

The above mentioned framework of spatial games with migration
specifies a kind of cellular automata (CA) model, in which space and
time are discrete. Furthermore, the set of actions performed in CA
models usually depends on the last time step only, which allows for
high-performance computing of large-scale scenarios. CA models have
been successfully applied to describe a variety of social and
physical systems \cite{ca0,ca1,ca2}. The outcomes of social and
physical interactions can be very well integrated into them, which
can create many interesting dynamics, like the ``game of life"
\cite{game_life}.

\section{Simulation Results and Discussion}

In the following, we perform a random sequential update of the
individual (i.e. their migration and learning steps), which is to be
distinguished from the parallel update assumed in
\cite{spatial_chaos,dis_env}. For a discussion of some related
advantages, see \cite{evo_game_sim}. Our numerical simulations are
performed on 49$\times$49 grids with $40\%$ empty sites, i.e. a
density of 0.6 and periodic boundary conditions.49x49 grids were
chosen for better visibility, while the statistical evaluations (see
Figs.~\ref{fig_phase_diag} and~\ref{fig_time_evo_noise}) were done
with 99$\times$99 grids for comparability with Nowak's and May's
spatial games \cite{spatial_chaos}. In the prisoner's dilemma, the
color code is chosen as follows: blue = cooperator, red = defector,
green = defector who became a cooperator, yellow = cooperator who
turned into a defector in the last iteration. In other games, blue =
player of group 1, red = player of group 2, green = player of group
2 who became a player of group 1, yellow = player of group 1 who
turned into a player of group 2. White always corresponds to an
empty site.

\subsection{Spontaneous Pattern Formation and Population Structure}
%

\begin{figure}[!htbp]
\begin{center}
\includegraphics[width=12cm, angle = 0]{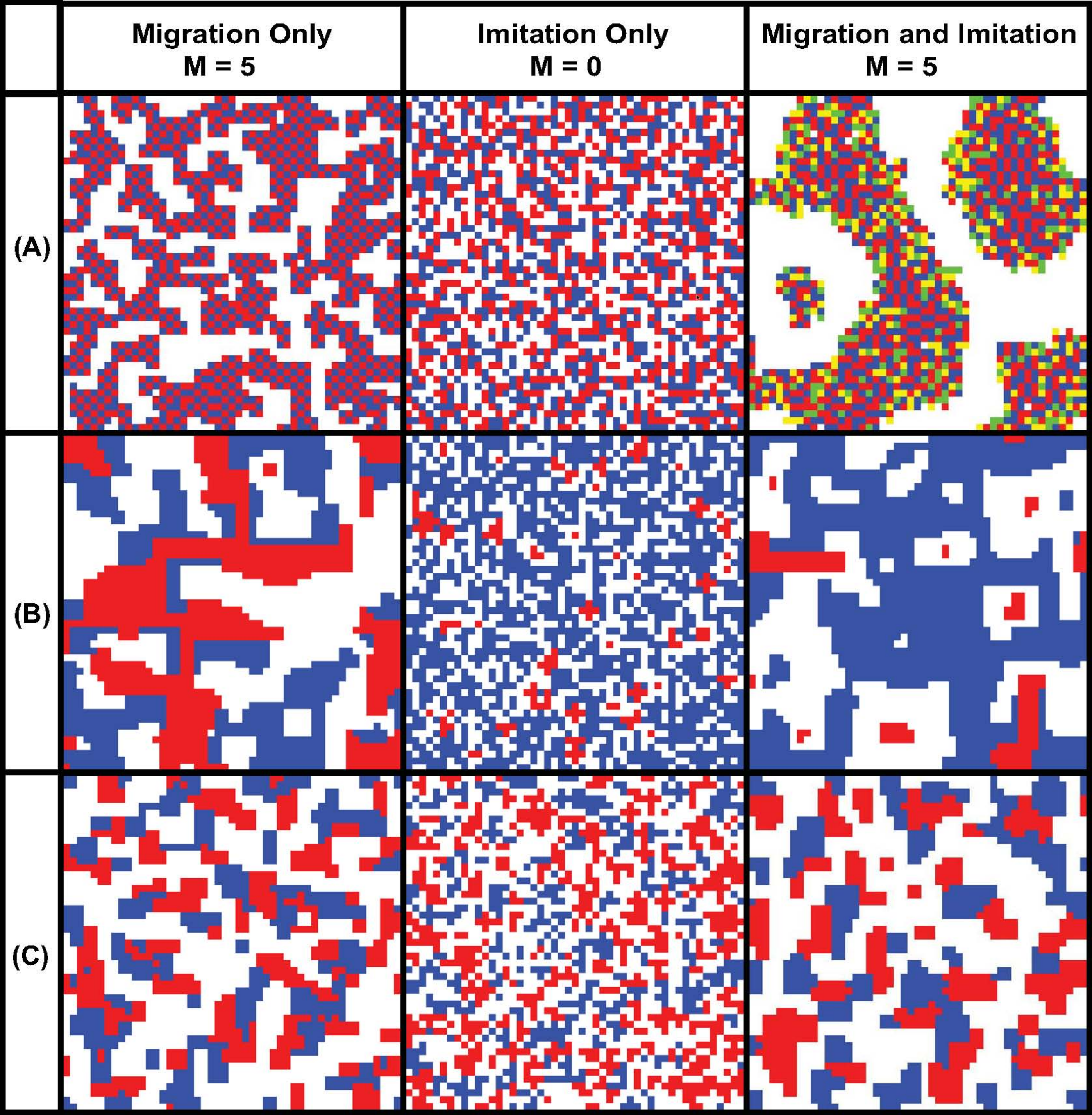}\,
\caption[]{ Simulation results for 49$\times$49 grids with a density
of 0.6 for the migration-only case (left), the imitation-only case
of conventional spatial games (middle) and the combination of
imitation with migration (right). The color code is chosen as
folows: blue = player of group 1, red = player of group 2, yellow =
player of group 1 who turned into a player of group 2 in the last
iteration, green = player of group 2 who turned into a player of
group 1 in the last iteration, white = empty site. \textbf{(A)}
$P_{11} = P_{22} = 0$, $P_{12} = P_{21} = 1$. Inter-group
interaction is encouraged, which causes the integration of two
populations. The combination of migration and imitation results in
the co-existence of both groups and the formation of clusters.
\textbf{(B)} $P_{11} = P_{22} = P_{12} = 1$, $P_{21} = -1$. When
affinity is unilateral, one population keeps approaching the other
one, which tends to evade. The combination of migration and
imitation leads to the spreading of the chasing group. Co-existence
is not observed. \textbf{(C)} $P_{11} = P_{22} = 1$, $P_{21} =
P_{21} = 0.5$. When interactions between groups are less profitable
than in the same group, residential segregation occurs as well, but
in contrast to \textbf{(B)}, it stabilizes.} \label{fig_pattern}
\end{center}
\end{figure}

In the migration game, individual preferences are reflected by the
payoff matrix, which quantifies the possible outcomes of social
interactions. Fig.~\ref{fig_pattern} shows a variety of patterns
formed when the payoff matrix is modified. The outcomes from
interactions between individuals are the driving forces changing the
population structure in space. In the following, we will call all
the individuals, who apply the same strategy, a ``group" $g$. Group
1 is represented in blue, group 2 in red. The size $N_{g}$ of group
$g$ is constant in true only in the migration-only case without
learning. Otherwise, strategy changes imply changes of individuals
between groups and changes in group size. We will distinguish the
three social relations between groups: (i) both groups (or, more
exactly speaking, their individuals) like each other $(P_{12} =
P_{21} = 1)$, (ii) group 1 is neutral with respect to group 2, but
group 2 dislikes group 1 $(P_{12} = 0, P_{21} = -1)$, and (iii)
intra-group interactions are more favored than inter-group
interactions ($P_{11} = P_{22} = 1$, $P_{12} = P_{21} = 0.5$).
Intra-group affiliation is always preferred or at least neutral
$(P_{11} = P_{22} = 1$ or $P_{11} = P_{22} = 0)$. As people are
allowed to move, when they like each other, a strong integration of
populations can be observed. However, if the affinity is unilateral,
then one population tends to evade the invasion of the other.
Assuming the same mobility in both populations, in the current
setting of our simulations the chasing population gains the majority
in the end, if unconditional imitation is considered.

Residential segregation emerges through the interactions of
individuals, not only when they dislike each other, but also if
individuals within the same group like each other more than
individual from other groups. When inter-group interactions result
in smaller payoffs, people attempt to maximize their payoff by
agglomerating with the same kind of people, which eventually results
in the segregation of different groups as a side effect.

\par

Some theoretical analysis can be useful to understand the
micro-macro link. Let $n_{g1}$ and $n_{g2}$ be the average number of
individuals of group 1 and 2, respectively, in the interaction
neighborhood of an individual using strategy $i$, i.e. belonging to
group $g=i$. Then, the payoff of an individual belonging to group
$g\in\{1,2\}$ is
\begin{eqnarray}
P_{g} = n_{g1}P_{g1} + n_{g2}P_{g2},
\end{eqnarray}
where $P_{gi}$ is the payoff of an individual of group $g$ when
meeting an individual using strategy $i$.

The total payoffs for individuals of group $g$ is $N_{g}P_{g}$,
while the total payoff of both groups is
\begin{eqnarray}
T'=N_{1}P_{1}+N_{2}P_{2},
\end{eqnarray}
where $N_{g}$ is the number of individuals of group $g$, i.e.
pursuing strategy $g=i$.



According to success-driven migration, the change of an individual's
payoff in the noiseless case is always positive, i.e.

\begin{eqnarray}
\Delta P_{g} = \Delta n_{g1}P_{g1} + \Delta n_{g2}P_{g2} > 0
\end{eqnarray}
with $\Delta n_{g1} = n_{g1}(t+1) - n_{g1}(t)$ and $\Delta n_{g2} =
n_{g2}(t+1) - n_{g2}(t)$.

Assuming $P_{12}\gg P_{11}$ and $P_{12} > 0$, we will usually have
$\Delta P_{1}\approx \Delta n_{12} P_{12} > 0$ and $\Delta n_{12} >
0$,
which leads to a monotonous increase in the number of neighbors of
individuals of group 1. Giving the inequality $P_{11}\gg P_{12}$ and
$P_{11}>0$, we analogously obtain $\Delta P_{1}\approx \Delta
n_{11}P_{11} > 0$ and $\Delta n_{11} > 0$. Therefore,  when
intra-group interactions of group 1 are much stronger than
inter-group interactions, and positive, clusters of group 1 will
expand, as each migration step will increase the average number of
neighbors within group 1. However, when $P_{11}<0$, the formation of
clusters of individuals belonging to group 1 by intra-group
interactions is unlikely. When $P_{12}<0$, an individual of group 1
tends to evade members of group 2, which can be regarded as a
repulsive effect attempting to keep a certain distance between
individuals of different groups \cite{panic}.

The total change of payoff is given by
\begin{eqnarray}
\Delta T' = N_{1} (\Delta n_{11}P_{11} + \Delta n_{12} P_{12}) +
N_{2} (\Delta n_{21}P_{21} + \Delta n_{22} P_{22}) > 0,
\end{eqnarray}
as long as success-driven migration is applied by all the
individuals, and changes $\Delta N_{1}$ and $\Delta N_{2}$ in the
group sizes are negligible.

Let us now consider a simple example, where $P_{11} = P_{22} = 0$.
Then, we have
\begin{eqnarray}
\Delta T' = N_{1}\Delta n_{12} P_{12} + N_{2}\Delta n_{21} P_{21},
\end{eqnarray}
which reflects the combined effects of interactions between members
of group 1 and 2. Furthermore, if $\Delta T' > 0$, we expect that
the number of neighborships between individuals of both groups will
monotonously increase. The corresponding macroscopic phenomenon is
the spatial integration (mixture) of both groups. However, if
$(P_{12} + P_{21}) < 0$, the reduction of inter-group links is
likely and will result in residential segregation (see
Fig.~\ref{fig_pattern}).

Here we have examined how individual preferences can change the
spatial population structure in a very simple social system
considering migratory behavior. The revealed social process can, to
some extent, reflect the dynamics of population succession in urban
areas. This corresponds to case \textbf{B} in
Fig.~\ref{fig_pattern}. Consider the payoff matrix in
Fig.~\ref{fig_pattern}\textbf{B}, and imagine that an individual of
group 1 happens to be located in the neighborhood of group 2. For
the previous residents, this may not change a lot. However, it
attracts other members of group 1, who are not yet living in a
neighborhood of group 1 individuals. This can trigger collective
migration of other group 1 members. Since interactions between
members of group 1 and 2 bring positive payoffs only for members of
group 1, group 2 will finally leave for new places. Of course, this
process repeats, just like the circle of population succession in
Sec. 1.2.

One may notice that, here, we do not differentiate the intention to
migrate and the actual migratory behavior. In daily life, however,
the movement of people is restricted by much more factors such as
wealth and time, so people do not migrate often, even if they are
motivated to move. But in our simulations as well, migration
activity after a few iterations is small: Starting with high
migration rate, due to the artificial choice of a random initial
distribution, the migration rate quickly drops to a low level
\cite{mig_game}. Of course, other factors determining migration can
be easily added to the above proposed framework of migration games.

\subsection{Promotion of Cooperation in the Prisoner's Dilemma}
The prisoner's dilemma is an important paradigm for studying the
emergence of cooperation among selfish individuals. Nevertheless,
studies of the prisoner's dilemma on lattices have not fully
explored the effect of mobility. It would be logical, of course, for
people to evade areas with a high level of defection, and to settle
down in areas dominated by cooperators. But would cooperative areas
be able to survive invasion attempts by defectors or even to spread
in space? To answer these questions, we will now focus on the
effects of success-driven migration on the spatial population
structure and the level of cooperation.

\begin{figure}[!htbp]
\begin{center}
\includegraphics[width=10cm, angle = 0]{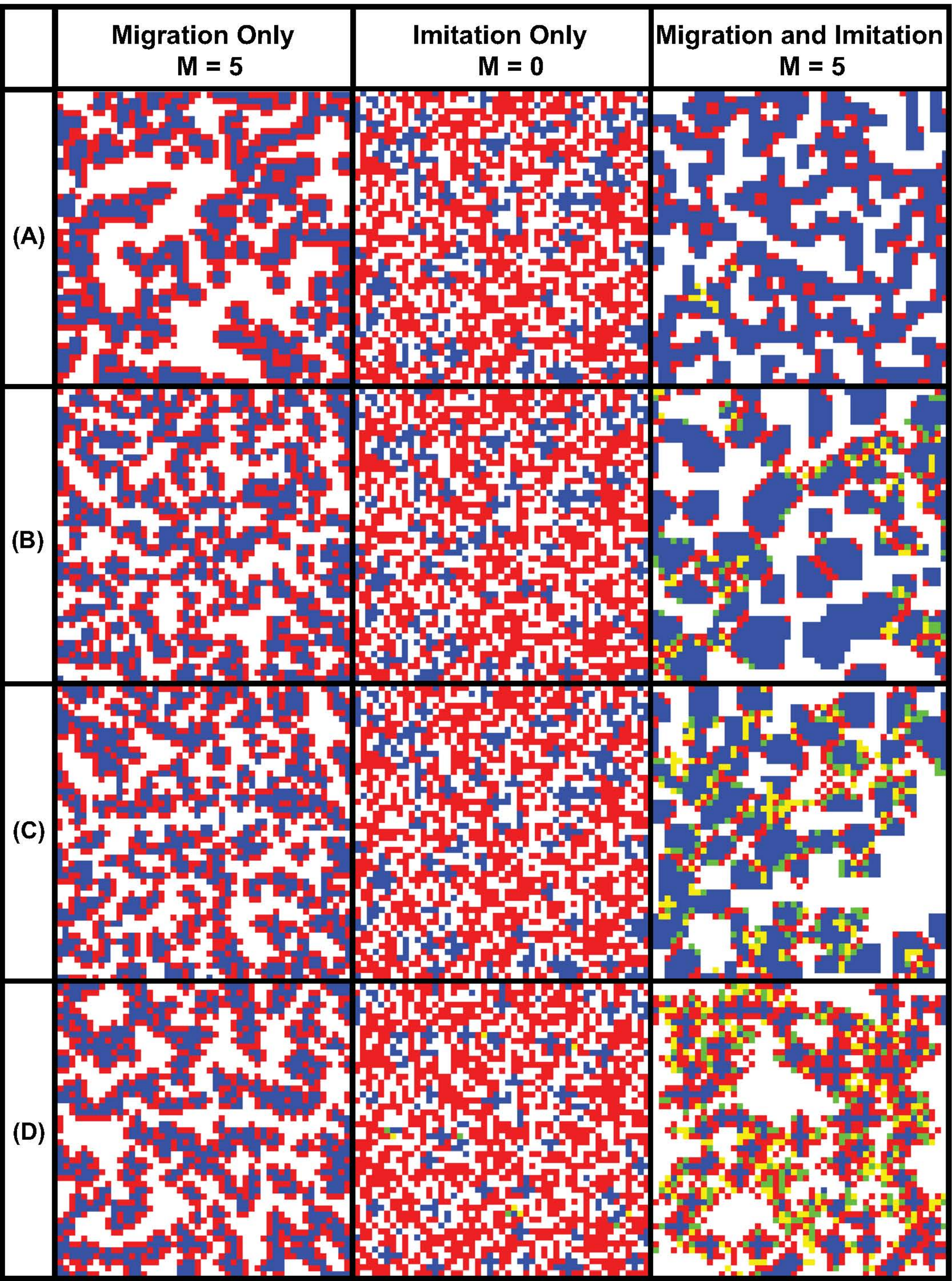}\,
\caption[]{Simulation results for 49$\times$49 grids with a density
of 0.6 for the migration-only case (left), the imitation-only case
of conventional spatial games (middle) and the combination of
imitation with migration (right). The color code is chosen as
follows: blue = player of group 1, red = player of group 2, yellow =
player of group 1 who turned into a player of group 2 in the last
iteration, green = player of group 2 who turned into a player of
group 1 in the last iteration, white = empty site. \textbf{(A)}
$P_{11} = R = 1$, $P_{12} = S = -0.2$, $P_{21} = T = 1.4$, $P_{22} =
P = 0$. The payoff matrix corresponds to a prisoner's dilemma.
\textbf{(B)} $P_{11} = R = 1$, $P_{12} = S = 0$, $P_{21} = T = 1.4$,
$P_{22} = P = 0$. The sucker's payoff is set to zero to be
compatible with the payoff matrix studied by Nowak and May.
\textbf{(C)} $P_{11} = R = 1$, $P_{12} = S = -0.2$, $P_{21} = T =
1.4$, $P_{22} = P = 0$. The migration and imitation step are
inverted here, i.e. an individual first imitates, then migrates.
\textbf{(D)} $P_{11} = 0.59$, $P_{12} = 0.18$, $P_{21} = 1$, $P_{22}
= 0$. In the snow-drift game, similar structures are found in the
migration-only and imitation-only case. Giving the possibility to
move, frequent switches of strategies are observed. See main text
for details.} \label{fig_pro_coop}
\end{center}
\end{figure}

Fig.~\ref{fig_pro_coop} compares the migration-only case with M = 5
(left) with the imitation only case corresponding to M = 0 (center)
and the combined imitation-and-migration case with M = 5 (right). In
the imitation-only case, the proportion of cooperators is greatly
reduced. However, the combination of migration and imitation
strikingly promotes the level of cooperation. Our explanation is
that, when individuals have mobility, cooperative clusters are more
likely to be promoted in the presence of invasion attempts of
defectors. We can see that, in the migration-only case, cooperators
manage to aggregate and to form clusters. Although defectors attempt
to enter cooperative clusters, they finally end up at the boundaries
of cooperative clusters, as cooperators split to evade defectors and
re-aggregate in new places, where defectors are excluded. In the
prisoner's dilemma, it is guaranteed that $2R > T + S$, which means
that the ``attractive force" between cooperators is mutual and
strong, while the interaction between a cooperator and a defector
leads to a unilateral attractive force $(T>0)$. When $S<0$, a
cooperator replies to defectors even in a repulsive way. Therefore,
defectors are less successful in joining or entering cooperative
clusters than cooperators are.

\par

\begin{figure}[!htbp]
\begin{center}
\includegraphics[width=8cm, angle = 0]{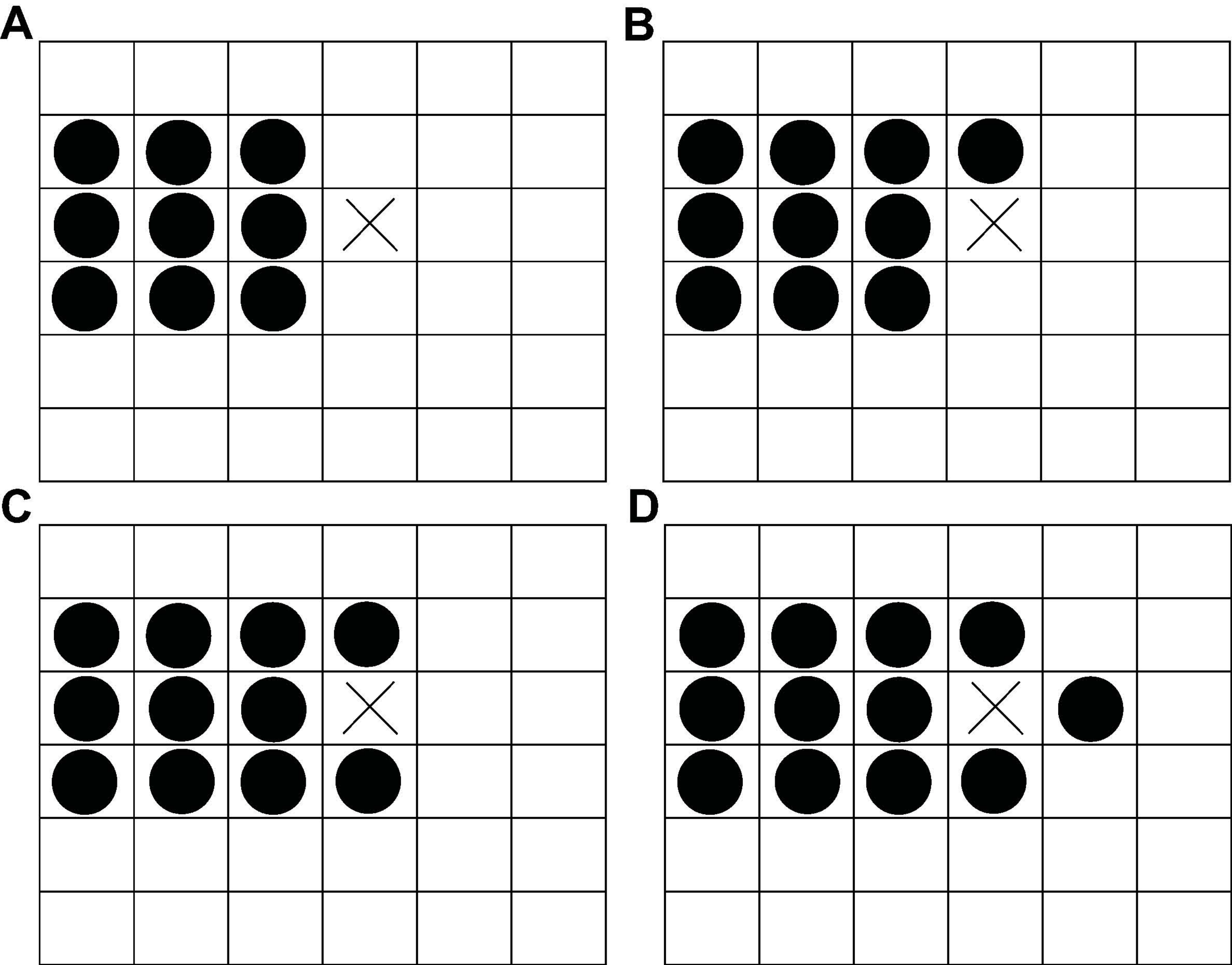}\,
\caption[]{ Analysis of the invasion of a cluster of cooperators
(black circles) by a single defector (cross). For the simplicity of
analysis, we assume $S = P = 0$ here, but one can easily generalize
the analysis to the situations with $P > S > 0$. \textbf{(A)}
Scenario 1: The payoff for the defector is $T$, while its
neighboring cooperator obtains a payoff of $3R$. Since $3R > T$, the
defector will become a cooperator. \textbf{(B)} Scenario 2: The
payoff for the defector is $2T$, while the maximal payoff among
neighboring cooperators is $3R$. In order to make the defector
become a cooperator, we need $3R > 2T$, i.e. $T < 1.5R$.
\textbf{(C)} Scenario 3: The defector obtains $3T$, while the
maximal payoff of neighboring cooperators is $3R$. In order to turn
the defector into a cooperator, the inequality $3R > 3T$ must be
satisfied, i.e. $T < R$. This condition can never be met in the
prisoner's dilemma. In order that cooperators do not copy the
defector, $4R > 3T$ must be satisfied, i.e. $T<\frac{4}{3}R$.
\textbf{(D)} Scenario 4: The payoff for the defector is $4T$, the
maximal payoff of cooperators in the whole community is $4R$.
Because of $T>R$, the defector can invade the cooperators nearby.
Once a cooperator becomes a defector however, the payoff for the
defectors will be reduced from $4T$ to $3T$, which may stop the
further invasion of defectors}\label{fig_config_analysis}
\end{center}
\end{figure}

Configurational analysis \cite{mig_game} (see
Fig.~\ref{fig_config_analysis}), indicates that, when $P = S = 0$,
and an individual only interacts with $k = 4$ nearest neighbors, a
cooperative cluster can turn a defector into cooperator by
unconditional imitation for $T < 1.5R$, when the defector is
surrounded with one or two cooperators. When a defector is
surrounded with three cooperators, cooperators can resist the
invasion of a defector, even if the temptation value $T$ is as high
as $\frac{4}{3}R$. A defector can invade one of the nearby
cooperators, if its neighborhood is fully occupied by cooperators.
Therefore, the formation of compact clusters is important to support
the spreading of cooperation. If $k>4$, the spreading of cooperation
occurs even for higher value of $T$.

In spatial games without mobility, the occurrence of a compact
cooperative clusters mainly depends on the initial distribution.
Giving mobilities, the migratory behavior can significantly
accelerate the formation of compact clusters, and promote the level
of cooperation.

\par

\begin{figure}[!htbp]
\begin{center}
\includegraphics[width=10cm, angle = 0]{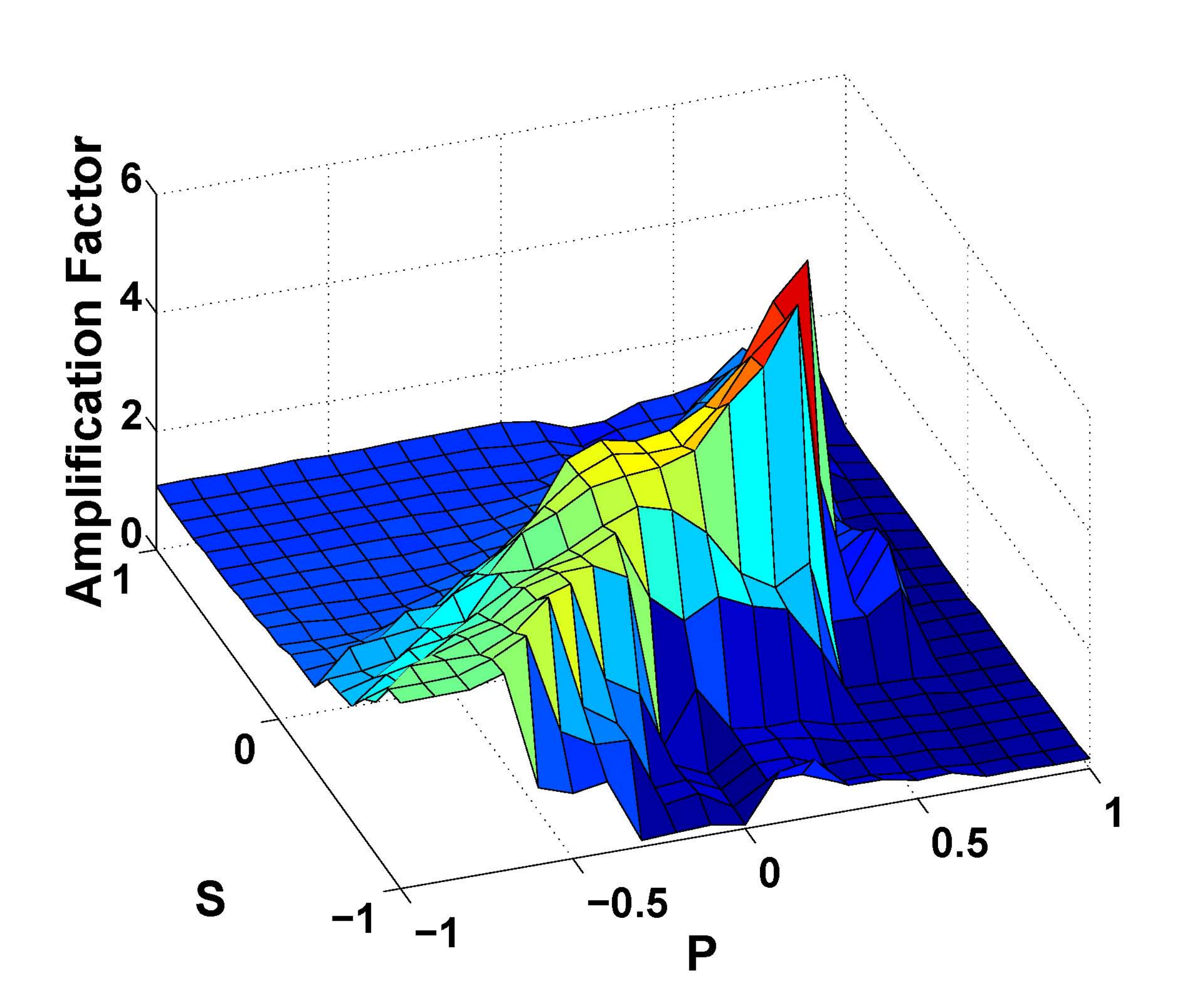}\,
\caption[]{Amplification factor of the level of cooperation by
migration ($M=5$) as a function of the sucker's payoff $P_{12}=S$
and the punishment $P_{22}=P$ in the prisoner's dilemma ($S<P$) and
the snow drift game ($S>P$). The simulation was performed for
99$\times$99 grids with a density of 0.6. $P$ and $S$ were varied
between $-1$ and 1, the payoffs $R = 1$ and $T = 1.3$ were left fix.
} \label{fig_phase_diag}
\end{center}
\end{figure}

Quantitative studies of how migration can promote the level of
cooperation have to compare the fraction of cooperators in
situations with mobility and without. Fig.~\ref{fig_phase_diag}
shows the amplification factor, defined as
\begin{eqnarray}
\delta(T,M) = \frac{f^{T}_{M}}{f^{T}_{0}}.
\end{eqnarray}
$f^{T}_{M}$ is the fraction of cooperators, given the mobility range
$M$ and a temptation value of $T$. We can see that the level of
cooperation in the prisoner's dilemma is promoted in a large
parameter area, if the punishment $P$ and the sucker's payoff are
roughly comparable in size.

\par

However, we have not yet studied the robustness of the migration
mechanism so far. One may imagine that, once a defector would manage
to invade a cooperative cluster, it may turn neighboring cooperators
into defectors as well, thereby eliminating cooperation eventually.
While in a noiseless environment, a defector cannot enter the center
of a compact cooperative cluster, ``noise" could make it happen with
a certain probability. As defectors in a cooperative cluster can
spread (see Fig.~\ref{fig_config_analysis}~\textbf{D}), noise could
therefore be thought to destroy the enhancement of cooperation by
success-driven migration. Very surprisingly, this is not the case!

In order to verify that success-driven migration robustly promotes
cooperation, we have implemented 3 kinds of noises. In each time
step, a certain proportion $y$ of individuals was selected to
perform
the following operations after the respective migration and imitation steps:\\
Noise 1: The selected players' locations were exchanged with a randomly chosen neighboring site (``neighborhood flipping").\\
Noise 2: The selected players' strategies were flipped, i.e. cooperation was replaced by defection and vice versa (``strategy flipping").\\
Noise 3: The selected players were removed from the grid, and an
equal number of players was created at randomly chosen free sites,
in order to mimic birth and death processes. The newly born players
had a $50\%$ chance to be cooperators and a $50\%$ chance to be
defectors.

\begin{figure}[!htbp]
\begin{center}
\includegraphics[width=12cm, angle = 0]{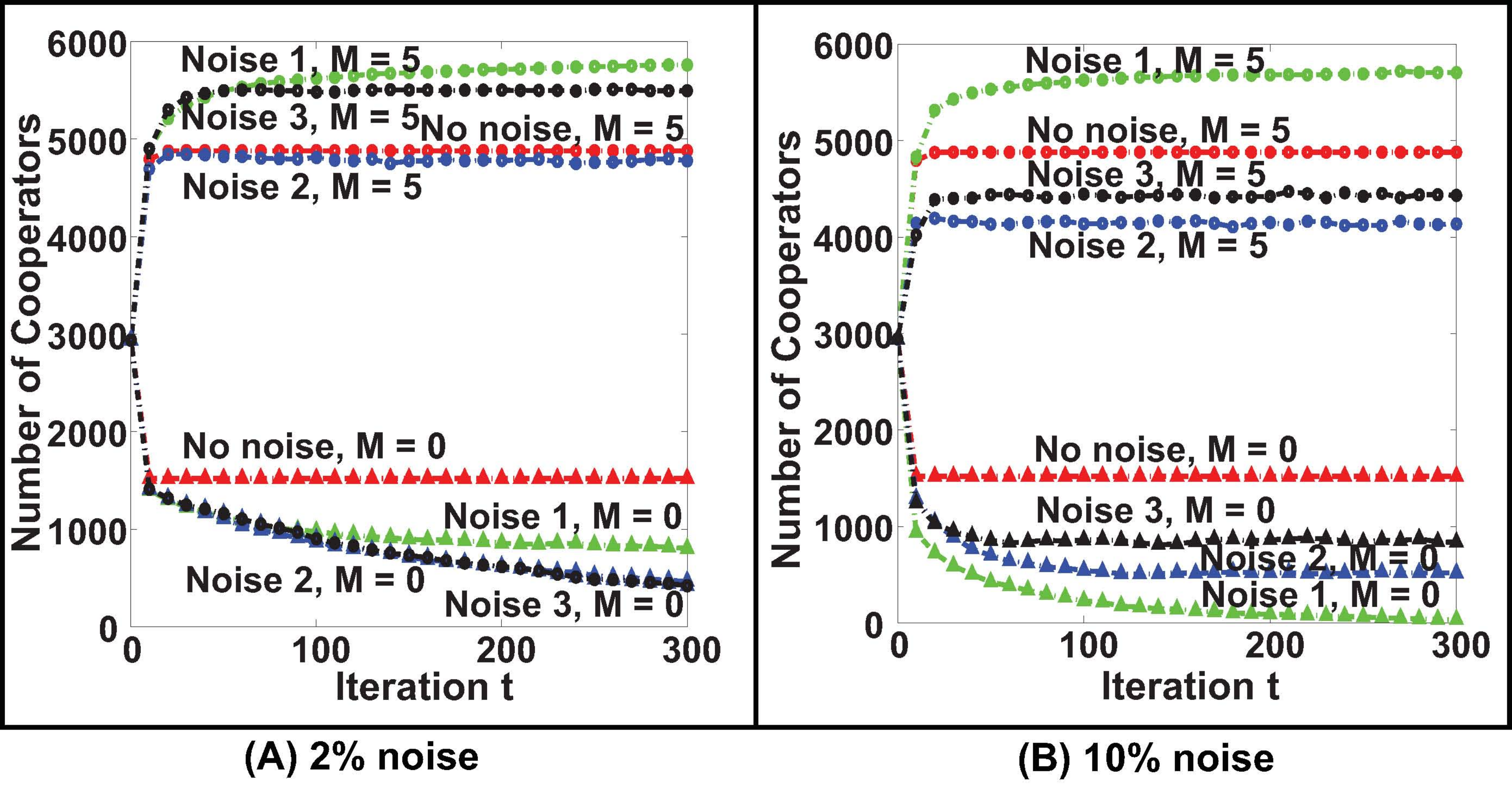}\,
\caption[]{Time evolution of the number of cooperators. Here, the
simulation is performed on 99$\times$99 grids with a density of 0.6
and payoffs $P_{11} = R = 1$, $P_{12} = S = 0$, $P_{21} = T = 1.3$,
and $P_{22} = P = 0$.} \label{fig_time_evo_noise}
\end{center}
\end{figure}

Fig.~\ref{fig_time_evo_noise} shows the time evolution of the number
of cooperators with noise strength $y=2\%$ and $y=10\%$
respectively. Without mobility, noise reduces the number of
cooperators greatly. For a mobility range $M=5$, however, we
surprisingly find that the level of cooperation can be still
maintained at a high level. With $2\%$ noise, noises 1 and 3 can
even increase the number of cooperators compared to the no-noise
case!

\par

\begin{figure}[!htbp]
\begin{center}
\includegraphics[width=4cm, angle = 0]{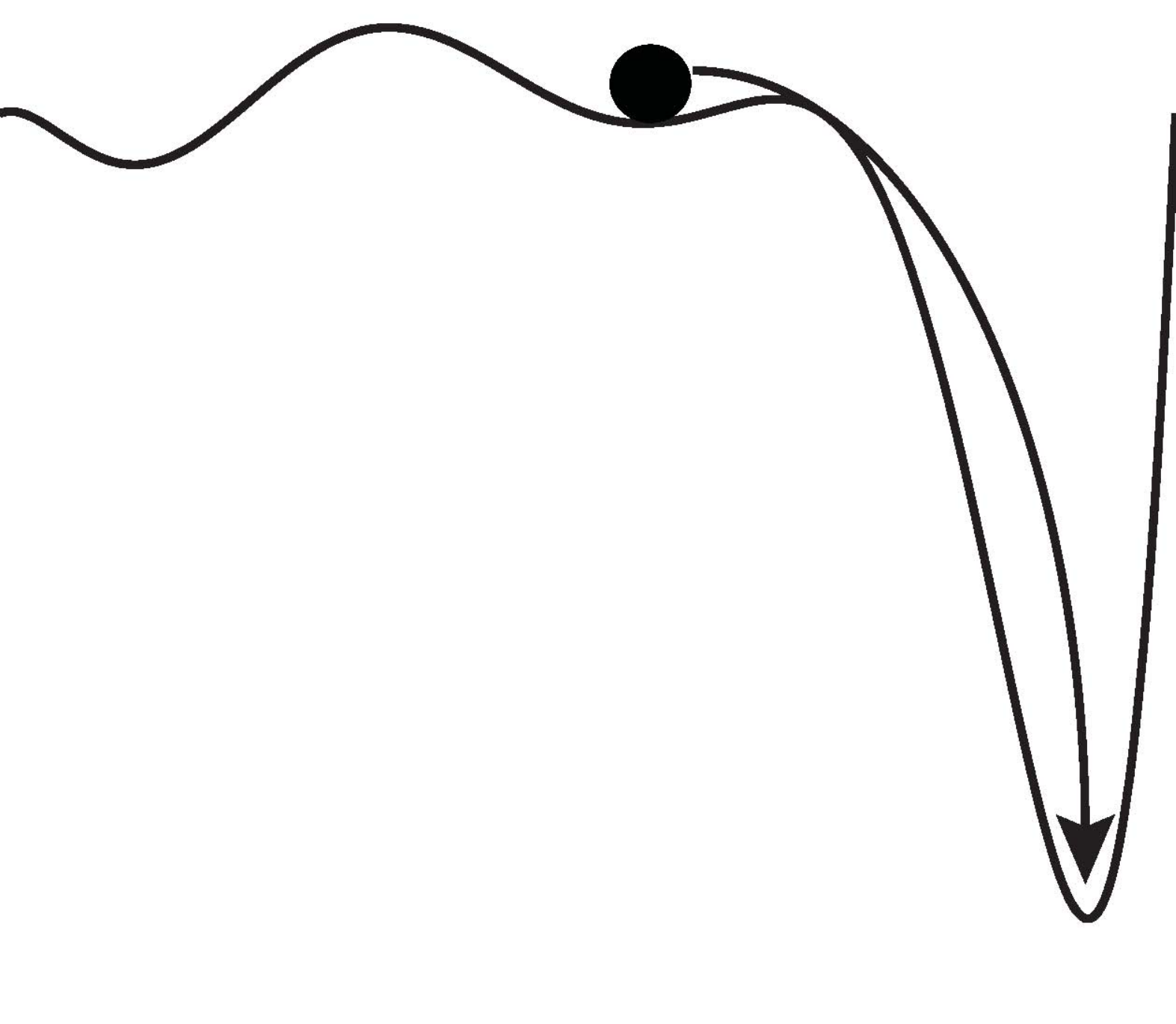}\,
\caption[]{Noise can make the system leave a sub-optimal state and
reach the globally optimal state.} \label{fig_hamilton_func}
\end{center}
\end{figure}

In order to understand how noise can promote cooperation to a level
higher than the level in a noiseless system, we may define a kind of
potential energy function of the system by the negative total
payoff:
\begin{eqnarray}
E = -T'.
\end{eqnarray}
In a noiseless system, each individual's migration step will
increase $T$, and reduce the potential energy of the system, [see
Eq. (11)]. Then an individual will adopt the most successful
strategy within its neighborhood. The imitation step and noise can
flip the strategies and increase the energy of the system. Just as
for the energy functions in spin glass models \cite{spin_glass},
recurrent neural networks \cite{recurrent_neuro} and Boltzman
machines \cite{boltz_mach}, there are many local minima of $E$ in
the migration game, one of which is reached within a few iteration
steps. Then the noiseless system behaves stationary. It is stuck in
the corresponding meta-stable state, which is highly dependent on
the initial condition. When noise occurs, it can perturb the
meta-stable state of the system and drive it to more stable states
(see Fig.~\ref{fig_hamilton_func}). Only if the noise level is too
strong, the system behaves pretty much in a random way. Therefore,
in conclusion, moderate levels of noise do not destroy a high level
of cooperation---they can even support it.

\par

\begin{figure}[!htbp]
\begin{center}
\includegraphics[width=12cm, angle = 0]{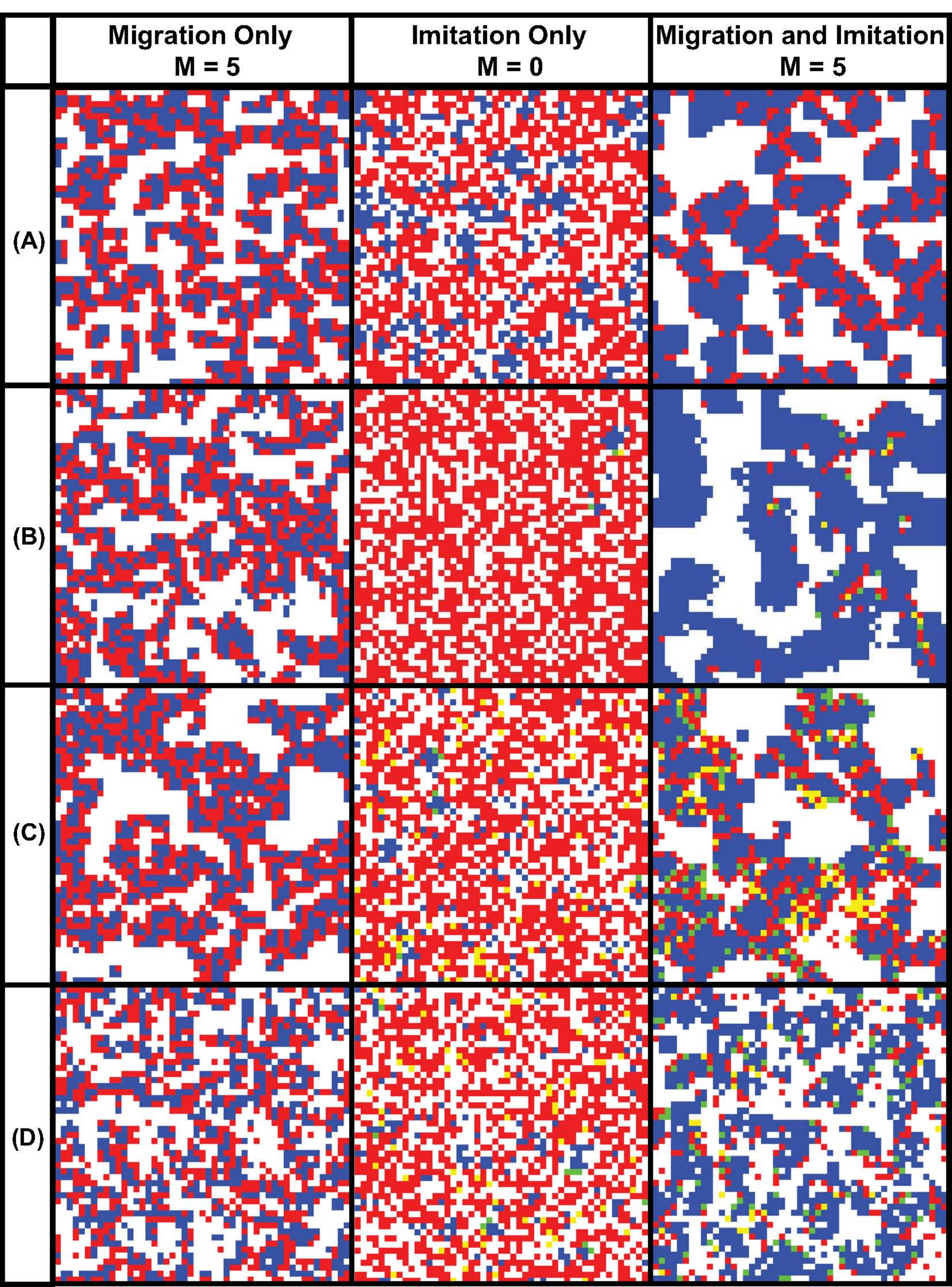}\,
\caption[]{Effect of different kinds of noises on the outcome of the
spatial prisoner's dilemma with migration but no imitation (left),
with imitation, but no migration (center), and with both, migration
and imitation (right). The simulation is performed on 49$\times$49
grids with a density of 0.6. The color code is chosen as follows:
red = defector, blue = cooperator, green = defector who became a
cooperator, yellow = cooperator who turned into a defector in the
last iteration, white = empty site. While the resulting level of
cooperation is very small in the conventional imitation-only case
(center), the additional consideration of migration results in large
levels of cooperation even in the presence of different kinds of
noise. \textbf{(A)} No noise. \textbf{(B)} Noise 1 (neighborhood
flipping). \textbf{(C)} Noise 2 (strategy flipping). \textbf{(D)}
Noise 3 (birth and death). The payoffs were $T = 1.3$, $R = 1$, $P =
0.1$, and $S = 0$ in all cases. See main text for details.}
\label{fig_pattern_noise}
\end{center}
\end{figure}

Our results for the conventional prisoner's dilemma with $P>S$
confirm the robustness of migration as a mechanism to promote
cooperation (see Fig.~\ref{fig_pattern_noise}). One can easily see
that, without mobility, the proportion of cooperators is close to
zero, while there is still a certain small number of cooperators in
the environment without noise. However, when individuals can move,
cooperation is significantly increased due to the spontaneous
formation of clusters. Imagine that a defector is located in the
center of a cooperators' cluster. In the beginning, defection can
invade cooperation due to the higher payoff (see
Fig.~\ref{fig_config_analysis}~\textbf{D}). But once a neighboring
cooperator becomes a defector, a defector's payoff is reduced from
$4T$ to $3T$, if $P = 0$. If more cooperators turn into defectors,
the payoff will be further reduced. Therefore, the exploitation of
cooperators by defectors is self-inhibitory, if individuals copy
better performing neighbors. Furthermore, a splitting of a
cooperative cluster can occur, since cooperators try to move to more
favorable neighborhood as soon as defectors are approaching them.
The migration of those cooperators can encourage other cooperators,
whose payoff depends on the mutual interactions, to move as well.
That is, defectors may trigger a collective movement of cooperators.
Of course, the newly formed cooperative clusters are also likely to
be invaded by neighborhood or strategy flipping, or birth and death
process. However, this will just repeat the above mentioned
migration process. Therefore, cooperators can survive and spread
even in a noisy world. Such a dynamical change of the population
structure through invasion and succession reflects various features
of the migratory behavior observed in reality (see sec. 1.2).

\par

\begin{figure}[!htbp]
\includegraphics[width=12cm, angle = 0]{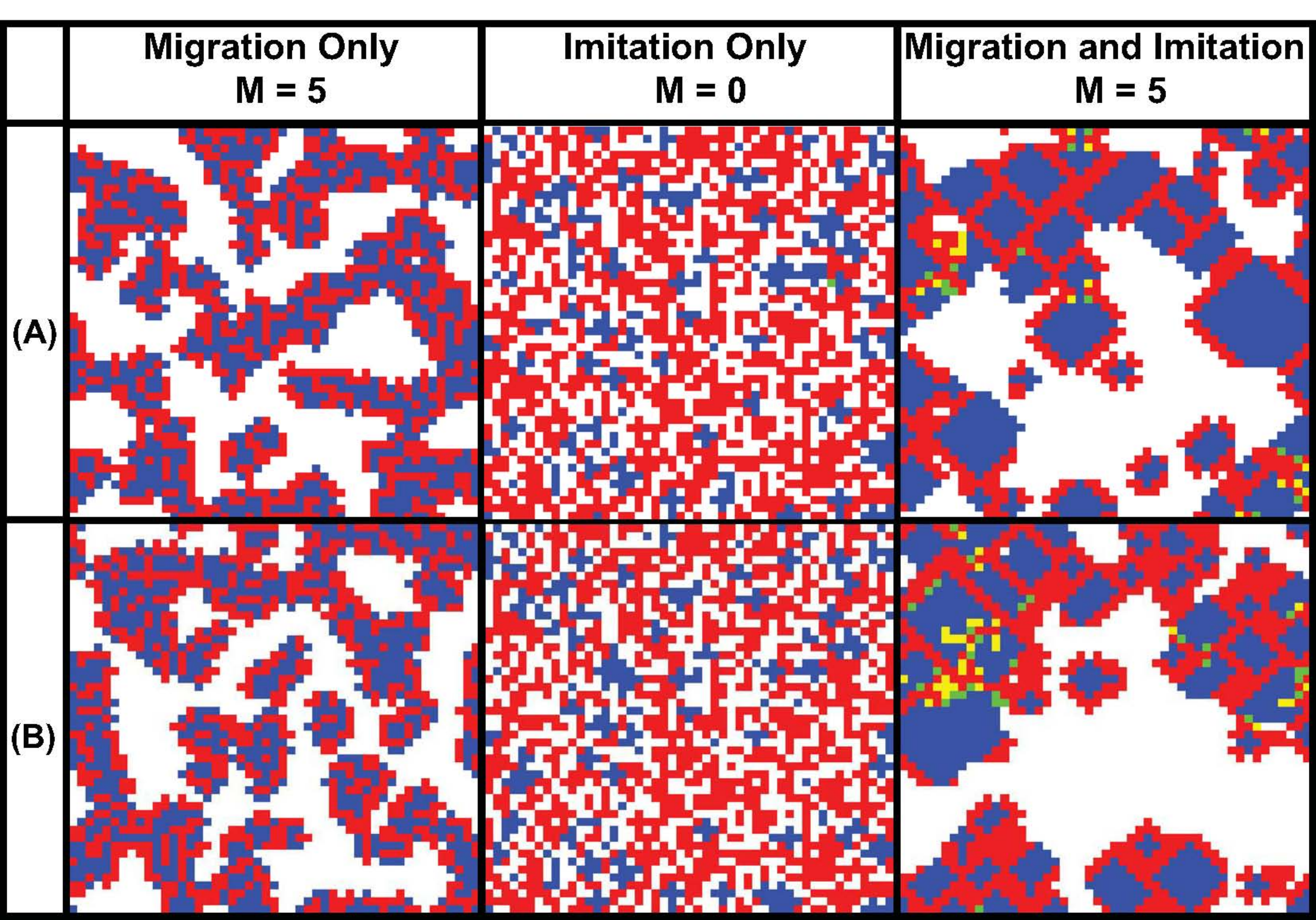}\,
\begin{center}
\caption[]{Migratory prisoner's dilemma with wealth-based
neighborhood-tagging rather than neighborhood testing as before. The
simulation is performed on 49$\times$49 grids with a density of 0.6.
The color code is chosen as follows: red = defector, blue =
cooperator, green = defector who became a cooperator, yellow =
cooperator who turned into a defector in the last iteration, white =
empty site. \textbf{(A)} $P_{11} = R = 1$, $P_{12} = S = 0$, $P_{21}
= T = 1.4$, $P_{22} = P = 0$. \textbf{(B)} $P_{11} = R = 1$, $P_{12}
= S = 0$, $P_{21} = T = 1.4$, $P_{22} = P = 0$, as in \textbf{(A)},
but the update rule is inverted, i.e. an individual, first imitates,
then migrates. In both cases, one can see that, without mobility,
the proportion of cooperators becomes very low. However, when
success-driven migration is possible, cooperation can spread, even
if the sequence of the migration step and the unconditional
imitation step is inverted.} \label{fig_fig3}
\end{center}
\end{figure}

In the migration rule studied above, we assume that a favorable
neighborhood can be determined by fictitious play, i.e. some
low-cost interactions with the people in that neighborhood
(``neighborhood testing"). One may think that this is difficult in
reality as it requires to reveal people's strategies. However, one
may also argue that people tend to migrate to high-quality
residential areas, which provide better education for children, a
low crime rate, and other social welfare. In fact, neighborhoods are
often ``labeled", and it may be assumed that this label (the
appearance and character of a neighborhood) depends on the total
cumulative payoffs (the accumulated wealth) of the residents in the
neighborhood. Therefore, one could assume that individuals try to
move to the neighborhood with the highest cumulative payoff. The
success-driven migration based on such a wealth-based ``neighborhood
tagging" is examined in Fig.~\ref{fig_fig3}. Again, we find that
migration promotes the formation of cooperative clusters and an
enhanced level of cooperation.

\par

\section{Conclusions}
We have introduced the concept of migration games by considering
success-driven motion. Migration games can easily reproduce
macroscopic stylized facts of various social phenomena based on
individual actions and interactions. Typical examples are population
succession and residential segregation. These aggregate outcomes
emerge from the interactions between individuals in a non-trivial
way, and a theoretical analysis allows one to qualitatively
understand the relation between the microscopic interactions and the
emerging macroscopic phenomena. Nevertheless, further studies are
required to fully elaborate the micro-macro link in a quantitative
way.

For the prisoner's dilemma, we have shown that self-organized
cooperative structures can promote the level of cooperation.
Moreover, we have verified that the enhancement of cooperation by
success-driven motion is robust to different kinds of noise.
Surprisingly, we even find that moderate noise levels can promote
the cooperation level further. The underlying mechanism is that, in
the migration game, success-driven motion will monotonously increase
the total payoff in the noiseless system, which however can lead the
system into a locally optimal state. The effect of noise can drive
the system out of local optima towards the globally optimal state.

The framework of migration games is quite flexible and allows the
integration of other interactions to study further social processes.

The spatial structure in our simulation is very simple, and the
neighborhood depends only on an individual's position. Real cities
are more complex and show a co-evolutionary dynamics. For example,
the city structure can reflect the distribution of social status
groups. Early models like the Burgess concentric zone model
\cite{burgess} divides the city into specific areas separated by
status rings. The city center is located in the middle of the
circle, around which newer, higher-quality housing stocks tend to
emerge at the perimeter of the city. Therefore, the growth of the
city center will expand adjacent residential zones outwards.

Status segregation is quite obvious in such an idealized model of
city structure. Poor people or new immigrants may only afford low
quality housing. Middle class people live in less compacted
neighborhoods. Rich people tend to accumulate in particular
quarters.

Burgess's model is based on the bid rent curve. Recognizing that
some poor people prefer to live near the main transportation
arteries and commercial establishments, Hoyt \cite{hoyt} modified
the concentric zone model to take this into account. In Hoyt's
model, cities tend to grow in wedge-shaped patterns or sectors.
Major transportation routes are emanating from the central business
district (CBD). The residential areas for lower income people are
located adjacent to the industrial quarters, while upper class
neighborhoods are far away from industrial pollution and noise.


It would be natural to extend migration games in order to study the
co-evolutionary dynamics of population structure and urban growth.
On the long run, we hope this will contribute to a better
understanding and planning of the population dynamics in a city.

Further research work can also study conflicts related with
migratory behavior, as has been revealed by the empirical research
\cite{conflicts_pattern,war_view}.


\begin{thebibliography}{99}
\bibitem{soc_context_seg}
Clark, W. A. and Fossett, M. Understanding the social context of the
Schelling segregation model. {\it Proc. Natl. Acad. Sci.} 105, 4109
- 4114 (2008).

\bibitem{ethnic_model}
Fossett, M. Ethnic preferences, social distance dynamics, and
residential segregation: theoretical explorations using simulation
analysis. {\it Journal of Mathematical Sociology} 30, 185-274
(2006).

\bibitem{social_dist0}
Duncan, O. D. and Duncan, B. Residential distribution and
occupational stratefication. {\it American Journal of Sociology} 60,
493-503 (1955).

\bibitem{social_dist1}
Reardon, S. F. and Firebaugh, G. Response: Segregation and social
distance - a generalized approach to segregation measurement. {\it
Sociological Methodology} 32, 85-101 (2002).

\bibitem{dyna_model_segre}
Schelling, T. C. Dynamic models of segregation. {\it J. Math.
Socio.} 1, 143-186 (1971).

\bibitem{micro_macro}
Schelling, T. C. {\it Micromotives and Macrobehavior} (W. W. Norton,
New York, 1978).

\bibitem{seg_poverty}
Massey, D. S. American apartheid: segregation and the making of the
underclass. {\it American Journal of Sociology} 96, 329-357 (1990).

\bibitem{closed_doors}
Yinger, J. {\it Closed Doors, Opportunities Lost: The Continuing
Costs of Housing Discrimination} (Russell Sage Found, New York,
1995).

\bibitem{ethnic_theo}
Macy, M., Rijt, A. V. D. Ethnic preferences and residential
segregation: Theoretical explorations beyond Detroit. {\it Journal
of Mathematical Sociology} 30, 275-288 (2006).

\bibitem{success_driven}
Helbing, D. and Platkowski T. Drift- or fluctuation-induced ordering
and self-organization in driven many-particle systems. {\it
Europhys. Lett} 60, 227-233 (2002).

\bibitem{opt_slef_org}
Helbing, D. and Vicsek, T. Optimal self-organization. {\it New
Journal of Physics} 1, 13.1-13.17 (1999).

\bibitem{mig_game}
Helbing, D. and Yu, W. Migration as a mechanism to promote
cooperation. {\it Advances in Complex Systems} 11, 641-652 (2008).

\bibitem{pop_success}
Cressey, P. F. Population succession in Chicago: 1898-1930. {\it
American Journal of Sociology} 44, 59-69 (1938).

\bibitem{five_rules_for_the_evolution_of_cooperation}
Nowak, M. A. Five rules for the evolution of cooperation. {\it
Science} 314, 1560 (2006).

\bibitem{the_evolution_of_cooperation}
Axelrod, A. {\it The Evolution of Cooperation} (Basic Books, New
York, 1984).

\bibitem{coop_comp_pathogenic_bacteria}
Griffin, A. S., West, S. A. and Buckling, A. Cooperation and
competition in pathogenic bacteria. {\it Nature} 430, 1024-1027
(2004).

\bibitem{neumann_game}
Von Neumann, J. and Morgenstern, O. {\it The Theory of Games and
Economic Behavior} (Princeton University Press, Princeton, 1944).

\bibitem{game_fuden}
Fudenberg, D. and Tirole, J. {\it Game Theory} (MIT Press,
Cambridge, 1991).

\bibitem{game_andreas}
Diekmann, A. Volunteer's dilemma. {\it Journal of Conflict
Resolution} 29, 605-610 (1985).

\bibitem{irregular_latt}
Flache, A. and Hegselmann, R. Do irregular grids make a difference?
Relaxing the spatial regularity assumption in cellular models of
social dynamics. In: {\it Journal of Artificial Societies and Social
Simulation} 4, no. 6 (2001), see
$http://www.soc.surrey.ac.uk/JASSS/4/4/6.html$.

\bibitem{replicator_dynamics}
Schuster, P. and Sigmund, K. Replicator dynamics. {\it J. Theor.
Biol.} 100, 533-538 (1983).

\bibitem{spatial_chaos}
Nowak, M. A. and May, R. M. Evolutionary games and spatial chaos.
{\it Nature} 359, 826-829 (1992).

\bibitem{spatial_structure_often_inhibits_the_evolition}
Hauert, C. and Doebell, M. Spatial structure often inhibits the
evolution of cooperation in the snowdrift game. {\it Nature} 428,
643-646 (2004).

\bibitem{human_ecology}
Park, R. E. Human ecology. {\it American Journal of Sociology} 42,
1-15 (1936).

\bibitem{neighbor_change}
Schwirian, K. P. Models of neighborhood change. {\it Ann. Rev.
Sociol.} 9, 83-102 (1983).

\bibitem{weidlich0}
Weidlich, W. Sociodynamics. {\it A Systematicc Approach to
Mathematical Modeling in the Social Sciences} (Harwoord Academic,
Amsterdam, 2000).

\bibitem{coll_behav}
Granovetter, M. Threshold models of collective behavior. {\it
American Journal of Sociology} 83, 1420-1443 (1978).

\bibitem{value_drop}
Harris, D. R. ``Property values drop when blacks move in,
because...": Racial and socioeconomic determinants of neighborhood
desirability, {\it American Sociological Review} 64, 461-479 (1999).

\bibitem{diff_game}
Hoogendoorn, S. and Bovy, P. H. L. Simulation of pedestrian flows by
optimal control and differential games. {\it Opim. Control Appl.
Mech.} 24, 153-172 (2003).

\bibitem{dis_env}
Vainstein, M. H. and Arenzon, J. J. Disordered environments in
spatial games. {\it Phys. Rev. E} 64, 051905 (2001).

\bibitem{mob_dec}
Vainstein, M. H., Silva, A. T. C. and Arenzon, J. J. Does mobility
decrease cooperation? {\it J. Theor. Biol.} 244, 722-728 (2006).

\bibitem{mob_visc}
Sicardi, E. A., Fort, H., Vainstein, M. H. and Arenzon, J. J. Random
mobility and spatial structure often enhance cooperation. {\it J.
Theor. Biol.} 256, 240-246 (2009).

\bibitem{ca0}
Epstein, J. M. {\it Generative Social Science} (Princeton University
Press, Prienceton, 2006).

\bibitem{indivi_stg}
Young, H. P. {\it Individual Strategy and Social Structure}
(Princeton University Press, Princeton, 1998).

\bibitem{learn_macy0}
Macy, M. W. Learning to cooperate: stochastic and tacit collusion in
social exchange. {\it American Journal of Sociology} 97, 808-843
(1991).

\bibitem{learn_macy1}
Macy, M. W. and Flache, A. Learning dynamics in social dilemmas.
{\it Proc. Natl. Acad. Sci.} 99, 7229-7236 (2002).

\bibitem{ca1}
Wolfram, S. {\it Theory and Applications of Cellular Automata}
(World Scientific Publication, Singapore, 1986).

\bibitem{ca2}
Chopard, B. and Droz, M. {\it Cellular Automata Modeling of Physical
Systems} (Cambridge University Press, Cambridge, 1998).

\bibitem{game_life}
Berlekamp, E. R., Conway, J. H. and Guy, R. K. {\it Winning Ways for
Your Mathematical Plays} (Academic Press, New York, 1982).

\bibitem{evo_game_sim}
Huberman, B. A. and Glance, N. S. Evolutionary games and computer
simulations. {\it Proc. Natl. Acad. Sci.} 90, 7716-7718 (1993).

\bibitem{panic}
Helbing, D., Farkas, I. and Vicsek, T. Simulating dynamical features
of escape panic. {\it Nature} 407, 487-490 (2000).

\bibitem{spin_glass}
Kirkpatrick, S., Sherrington, D. Infinite-ranged models of
spin-glasses. {\it Physical Review B} 17, 4384-4403 (1978).

\bibitem{recurrent_neuro}
Hopfield, J. J. Neural networks and physical systems with emergent
collective computational abilities. {\it Proc. Natl. Acad. Sci.} 79,
2554-2558 (1982).

\bibitem{boltz_mach}
Ackley, D. H., Hinton, G. E., Sejnowski, T. J. A learning algorithm
for Boltzmann machines. {\it Cognitive Science} 9, 147-169 (1985).

\bibitem{burgess}
Burgess, E. W. Residential segregation in American cities. {\it
Annals of the American Academy of Political and Social Science }
140, 105-115 (1928).

\bibitem{hoyt}
Hoyt, H. {\it The Structure and Growth of Residential Neighborhoods
in American Cities Washington} (Federal Housing Administration,
1939).

\bibitem{conflicts_pattern}
Lim, M., Metzler, R. and Yam, Y. B. Global pattern formation and
ethnic/cultural violence. {\it Science} 317, 1540-1544 (2007).

\bibitem{war_view}
Weidmann, N. B., and Kuse, D. WarViews: Visualizing and animating
geographic data on civil war. Accepted for publication in
International Studies Perspectives.

\end{thebibliography}
\end{document}